\documentclass[twocolumn,showpacs,preprintnumbers,amsmath,amssymb,prl]{revtex4}

\usepackage{graphicx}
\usepackage{dcolumn}
\usepackage{bm}

\begin{document}

\title{A clock transition for a future optical frequency standard
with trapped atoms}

\author{Ir\`{e}ne Courtillot}
\author{Audrey Quessada}
\author{Richard P. Kovacich}
\author{Anders Brusch}
\author{Dmitri Kolker}
\author{Jean-Jacques Zondy}
\author{Giovanni D. Rovera}
\author{Pierre Lemonde}

 \email{pierre.lemonde@obspm.fr}
\affiliation{BNM-SYRTE, Observatoire de Paris\\ 61, Avenue de
l'observatoire, 75014, Paris, France}

\date{\today}

\begin{abstract}
We report the first direct excitation of the strongly forbidden
$5s^{2}~^{1}S_0-5s5p~^3P_0$ transition in $^{87}$Sr. Its frequency
is 429 228 004 235 (20) kHz. A resonant laser creates a small leak
in a magneto-optical trap (MOT): atoms build up to the metastable
$^3P_0$ state and escape the trapping process, leading to a
detectable decrease in the MOT fluorescence. This line has a
natural width of $10^{-3}$ Hz and can be used for a new generation
of optical frequency standards using atoms trapped in a light
shift free dipole trap.
\end{abstract}

\maketitle

In addition to being at the heart of the international system of
units, extreme frequency metrology finds a wide range of
applications throughout physics. As an example, tests of the
stability of fundamental constants are performed by comparing the
frequencies of different atomic transitions
\cite{Marion03,Bize03}. These laboratory tests are already
competitive with those performed at the cosmological scale
\cite{Damour96,Webb01} and improve with the accuracy of the atomic
transitions frequency measurements. Microwave frequency standards
such as atomic fountains now have relative accuracies better than
$10^{-15}$ with potential improvements down to $10^{-16}$
\cite{Bize02}. Going further, however, seems extremely difficult
due to the limited quality factor of the atomic resonance in these
devices ($\sim 10^{10}$). Optical frequency standards exhibit much
higher line-$Q$, up to $10^{14}$ as recently demonstrated in Ref.\
\cite{Rafac00}. These devices reach fractional frequency
instabilities below $10^{-14}$ over 1 s \cite{Diddams01}, almost
one order of magnitude better than fountains \cite{Bize02}. Their
current accuracy lies near 1 part in $10^{14}$ with anticipated
large room for improvement
\cite{Niering00,Bize03,Riehle02,Curtis02,StengerOL01}.

Two different approaches are commonly used to develop optical
frequency standards \cite{StAndrews}. The first one is based on
the spectroscopy of a single trapped ion, the second on the
spectroscopy of a large ensemble of free falling neutral atoms. It
is a commonly shared opinion that the ion approach may lead to a
better ultimate frequency accuracy due to the "perfect" control of
the ion motion, while the atomic approach should lead to a better
frequency stability thanks to the numerous quantum references
contributing to the signal. Recently H. Katori proposed a scheme
which would combine the advantages of both approaches
\cite{Katori02}. The idea is to trap neutral atoms in the
Lamb-Dicke regime in an optical lattice operating at a wavelength
where the light-shift of the clock transition vanishes.

H. Katori proposed to use $^{87}$Sr probed on the
$5s^{2}~^{1}S_0-5s5p~^3P_0$ line at 698\,nm, which indeed seems an
ideal system for the realization of this scheme. This $J=0-J=0$
resonance is only slightly allowed by hyperfine coupling and its
metrological properties are exquisite. A natural linewidth of
$\sim 1$ mHz can be derived from the hyperfine data found in Ref.\
\cite{Kluge74}. It has a high insensitivity to external
electromagnetic fields. The light-shift cancellation is expected
to occur near 800\,nm, a wavelength far from any atomic resonance
and for which powerful and practical laser sources are readily
available. Apart from a residual effect due to hyperfine
structure, the light shift cancellation is independent on
polarization. Finally, higher order effects due to the trapping
field are expected to be extremely small on this particular line
\cite{Pal02}. In this Letter we report the first direct
observation and frequency measurement of this transition.

\begin{figure}[htb]
\begin{center}
\includegraphics[width = \columnwidth]{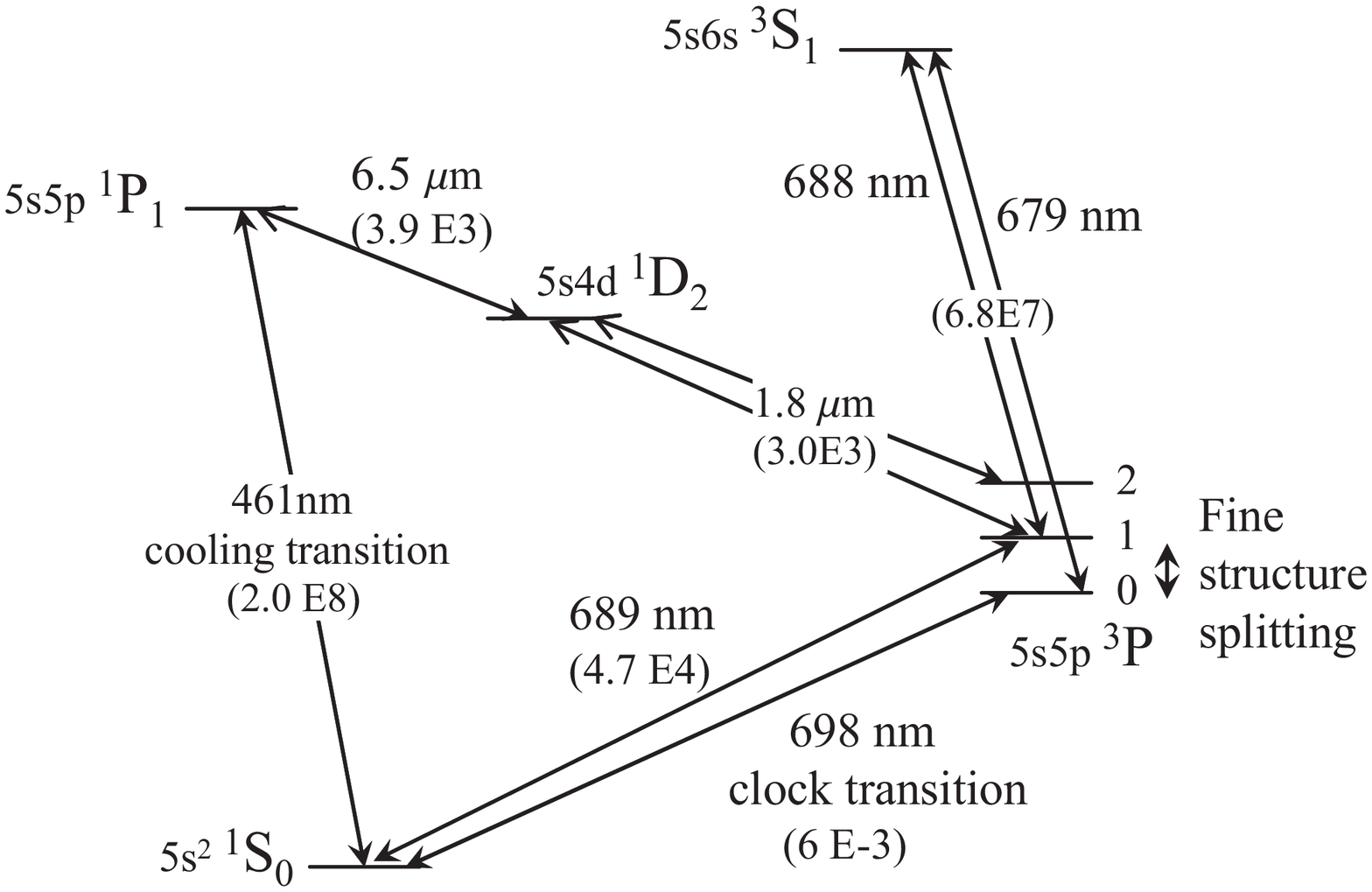}
\end{center}
\caption{Energy diagram of $^{87}$Sr with wavelength and decay
rate (s$^{-1}$) of the transitions involved in the experiment. For
clarity, the hyperfine structure is not represented ($I=9/2$).}
\label{fig:levels}
\end{figure}

The experiment is performed on a sample of cold atoms collected in
a magneto-optical trap (MOT). An atomic beam is decelerated in a
Zeeman slower and captured at the crossing-point of three
retro-reflected beams tuned 40 MHz to the red of the $^1S_0-
\,^1P_1(F=9/2)$ transition (Fig.\ \ref{fig:levels}). The setup is
described in details in Ref.\ \cite{Courtillot03}. For this
particular experiment $3\times 10^6$ $^{87}$Sr atoms are trapped
at steady state at a temperature of 2 mK. The atomic cloud has a
$1/e^2$ diameter of $2$ mm. To probe the 698 nm line, a 14 mW
laser beam passes four times through the atomic cloud in a
standing wave configuration. The probe beam waist radius is 1.3
mm. At first sight, a direct detection of the very forbidden line
with our system seems a desperate task: the resonance is expected
to be Doppler broadened to 1.5 MHz (FWHM) more than three orders
of magnitude higher than the Rabi frequency (0.8 kHz). Thus only
$10^3$ atoms are expected to be resonant with the laser at a time.
In addition, an efficient fluorescence detection of atoms in the
$^3P_0$ state is made difficult by the absence of cycling
transition from this state. Finally the energy of the $^3P_0$
state is known from databases to within a few hundred MHz only.

To overcome these problems we have first determined the resonance
frequency with an accuracy of 110 kHz by measuring the frequency
of the $^1S_0-\,^3P_1$ line at 689 nm and of the frequency
difference between the $^3P_1-\,^3S_1$ and $^3P_0-\,^3S_1$
transitions at 688\,nm and 679\,nm respectively (see Fig\
\ref{fig:levels}) \cite{Note}. Second, to amplify the fraction of
cold atoms transferred to the $^3P_0$ state by direct excitation,
we have taken advantage of the fact that the lifetime of the MOT
is 40 times larger than the Rabi oscillation period. A laser tuned
to resonance induces a leak in the MOT leading to a detectable
decrease of the number of trapped atoms of 1\%.

\begin{figure}[htb]
\begin{center}
\includegraphics[width = \columnwidth]{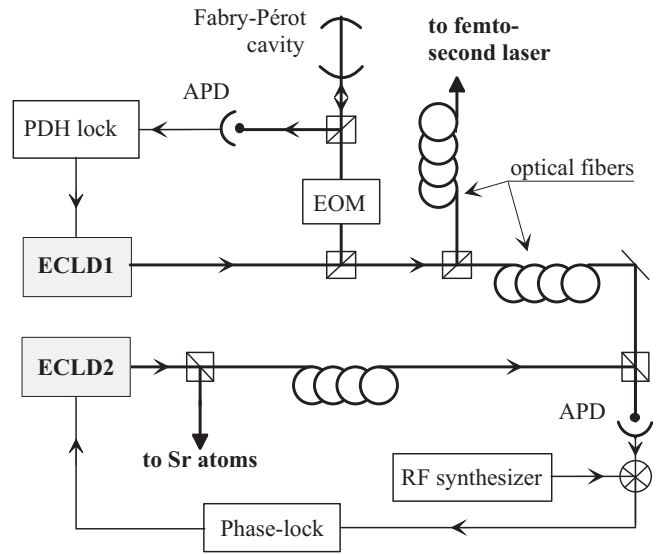}
\end{center}
\caption{Experimental set up used for frequency measurements. An
extended cavity laser diode (ECLD1) is locked to a high finesse
Fabry-P\'{e}rot cavity (F=27000). 100 $\mu$W are sent to a
frequency chain based on a femtosecond laser for absolute
frequency measurement. A second laser (ECLD2) which is used to
probe the strontium atoms is offset-phase locked to ECDL1. The
beat-note collected by an avalanche photodiode (APD) is mixed with
a radio frequency reference. Noise and frequency offsets
introduced by optical fibers are negligible in this experiment.
EOM: electro-optical modulator.} \label{fig:exp}
\end{figure}

The experimental set up used for all frequency measurements is
described in Fig.\ \ref{fig:exp}. An extended cavity laser diode
(ECLD1) is locked to a high finesse cavity using the
Pound-Drever-Hall method \cite{Drever83} with performances
reported in Ref.\ \cite{Quessada03}. Its frequency is continuously
measured vs a hydrogen maser with a scheme based on a self
referenced femtosecond Ti:Sapph laser \cite{Holzwarth00,Jones00}.
The relative resolution of this measurement is typically $3\times
10^{-13}$ for a one second averaging time. A second laser (ECLD2)
is offset-phase locked to ECLD1. The beat-note between both lasers
is mixed with the output of a radio frequency synthesizer to
generate the offset phase lock error signal. The bandwidth of the
servo control is 2 MHz. With this scheme the light of ECLD2, which
is sent to the atoms, can be tuned to any frequency between two
modes of the cavity of free spectral range 1.5 GHz by actuating
the RF synthesizer. Two sets of lasers are used. One can be tuned
from 675 to 685 nm, the other one from 685 to 698 nm.

The frequency of the $^1S_0-\,^3P_1(F=9/2)$ transition is measured
with an atomic beam independent from the cold atom setup. The
obtained frequency is 434 829 342 950 (100) kHz, with an
uncertainty mainly dominated by statistical noise and imperfect
knowledge of the magnetic field.

\begin{figure}[htb]
\begin{center}
\includegraphics[width=\columnwidth]{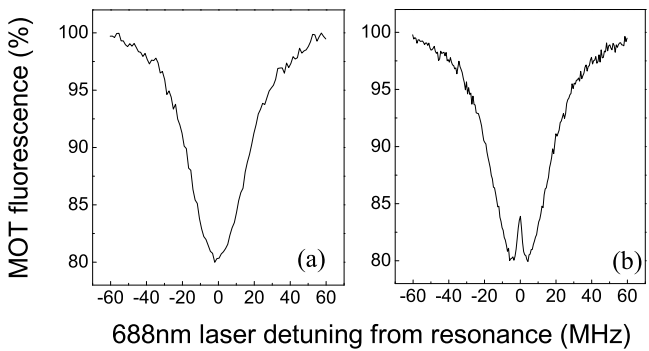}
\end{center}
\caption{Relative MOT fluorescence signals obtained when a 688\,nm
laser is swept around the $^3P_1,F=9/2-^3S_1,F=11/2$ transition.
(a) 688 nm laser only; (b) with an additional 679 nm laser tuned
to the $^3P_0,F=9/2-^3S_1,F=11/2$ resonance. In this case the CPT
dip is observed. Intensity of the 688 nm and 679 nm laser are 5
$\mu$W/mm$^2$ and 2 mW/mm$^2$ respectively.} \label{fig:CPT}
\end{figure}

The determination of the fine structure splitting between $^3P_0$
and $^3P_1$ is performed with the cold atoms. While trapped and
cycling on the $^1S_0-\,^1P_1$ transition, atoms eventually emit a
spontaneous photon which brings them to the $^1D_2$ state (Fig.\
\ref{fig:levels}). This state has two main decay channels: to the
$^3P_1$ and $^3P_2$ states. Atoms in the $^3P_1$ state decay back
to the ground state and are kept in the trap while atoms in the
metastable $^3P_2$ state are lost. This process limits the
lifetime of the MOT to some 50 ms. We make the fine structure
measurement by modifying this escape process. If a laser resonant
to one of the hyperfine components of the $^3P_1-\,^3S_1$
transition is added to the trap, atoms in the corresponding
$^3P_1$ state are pumped to the $^3P_2$ and $^3P_0$ metastable
states. They escape the trap instead of decaying back to the
ground state. This decreases the trapped atom number. Fig.\
\ref{fig:CPT}(a) shows the fluorescence of the MOT as a function
of the 688 nm laser detuning from the $^3P_1,F=9/2-\,^3S_1,F=11/2$
resonance.

A direct detection of the $^3P_0-\,^3S_1$ transition by the same
technique is not possible because $^3P_0$ is not populated in the
MOT. Instead, a first signal is obtained by detecting the light
shift of the $^3P_1-\,^3S_1$ transition induced by a 679 nm laser
close to the $^3P_0-\,^3S_1$ resonance. With an intensity of 1.8
mW/mm$^2$, a shift of 100 kHz is observed for a detuning of 100
MHz. For a 679 nm detuning smaller than the width of the 688 nm
resonance, a dip appears in the resonance profile due to coherent
population trapping (CPT): when the frequency difference between
both lasers matches the atomic fine structure, there exists a
coherent superposition of $^3P_1$ and $^3P_0$ states which is not
coupled to $^3S_1$ \cite{Alzetta76}. Atoms in this dark state are
not pumped to $^3P_2$ but decay back to the ground state in a few
10 $\mu$s due to the $^3P_1$ instability. They are kept in the
MOT. When the CPT dip is centered on the 688 nm resonance, the 679
nm laser is tuned to resonance. The observed signal in this
configuration is shown in Fig.\ \ref{fig:CPT}(b). The fine
structure measurement is performed light shift free with both
lasers locked on resonance. We measured a frequency of 5 601 338
650 (50) kHz with an uncertainty mainly due to the magnetic field
gradient of the MOT. According to these measurements, the
$^{1}S_0-\,^3P_0$ transition is expected to have a frequency of
429 228 004 300 (110) kHz.

For the direct observation, we induce in the MOT a leak to the
$^3P_0$ state with a laser tuned to resonance. The lifetime of the
trap is two orders of magnitude longer than the duration of a
$\pi$ pulse on the forbidden transition with the laser parameters
given above. This leads to a build-up by the same factor of the
fraction of atoms escaping the MOT if the transfer rate to $^3P_0$
is constant and if the atoms, once in the $^3P_0$ state, actually
escape the trapping process. One way to fulfil the first condition
could consist in using the MOT to rethermalize atoms to fill the
dip in the velocity distribution created by excitation to $^3P_0$.
For the escape condition, one could think of optical pumping to
$^3P_2$ with a laser tuned to $^{3}P_0-\,^3S_1$. In our experiment
however, the Doppler effect induced by gravity is sufficient to
fulfil both conditions. It amounts to several times the Rabi
frequency per millisecond with the $45^\circ$ angle formed by the
698\,nm probe beam and vertical. The dip in the velocity
distribution induced by the excitation laser is then permanently
refilled. On the other hand, atoms in the $^3P_0$ state are
rapidly detuned from the excitation laser and do not emit a
stimulated photon back to the ground state. The trapped atom
number should then decrease by several \%.

\begin{figure}[htb]
\includegraphics[width=\columnwidth]{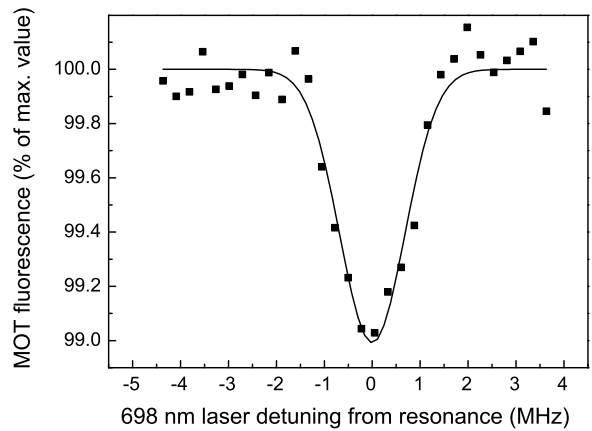}
\caption{Direct observation of $^{1}S_0-\,^3P_0$ transition. The
line is broadened by the Doppler effect due to the finite
temperature of the atoms to 1.4 MHz
(FWHM).} \label{fig:698} 
\end{figure}

In practice, the experiment is operated sequentially: by means of
acousto-optic modulators we alternate a capture and cooling phase
with the blue lasers and a probe phase with the 698 nm laser. The
main motivation for this pulsed operation is to avoid any light
shift of the forbidden transition. Another reason is related to
the escape process of the $^3P_0$ atoms. It would be less
efficient with the MOT beams on due to the instability of the
$^1S_0$ state confered by the coupling to $^1P_1$: the reduced
lifetime of the atomic coherences due to spontaneous emission of
blue photons would lead to an effective broadening of the 698\,nm
excitation much larger than the Doppler effect due to gravity. The
optimization of the time sequence results from a trade-off between
the capture efficiency of the MOT, the ballistic expansion of the
atomic cloud during probe phases, and the efficiency of excitation
to $^3P_0$. With phases of capture of 3 ms duration and probe
phases of 1 ms duration, we have $1\times 10^6$ atoms at steady
state in the MOT and the contrast of the resonance is 1\%. In
figure \ref{fig:698} is shown the fluorescence of the trapped
atoms versus the 698 nm laser detuning from resonance. A narrow
sub-Doppler structure is expected at the center of the resonance
due to the standing wave configuration. With the present signal to
noise ratio, we are not able to detect it.

We have locked the laser to the $^{1}S_0-^{3}P_0$ resonance. Its
frequency is 429 228 004 235 kHz with a standard deviation of 20
kHz for an averaging time of 2 hours. At that level, the
systematic effects are negligible. The first order Doppler effect
vanishes due to the standing wave configuration. The Zeeman effect
induced by the MOT field gradient is less than 1 kHz, since the
Land\'{e} factors of the $^{1}S_0$ and $^{3}P_0$ levels are both
equal to $-1.3\times 10^{-4}$. The light shift which would result
from imperfect extinction of the blue light is also estimated
below 1 kHz. Tab.\ \ref{tab:Freq} summarizes the indirect and
direct measurements and confirms the good agreement between both
methods.

\begin{table}[htb]
\caption{Measured frequencies of $^{87}$Sr transitions ($F$=9/2
for all atomic states).}
\begin{center}
\begin{tabular}{cl}
\hline\hline\textbf{Transition} & \multicolumn{1}{|c}{\textbf{Frequency (kHz)}}   \\
\hline\hline $5s^2\,^{1}S_0-5s5p\,^3P_1$ & \multicolumn{1}{|l} {434 829 342 950 $\pm$ 100}\\
\hline $5s5p\,^3P_0-5s5p\,^3P_1$ & \multicolumn{1}{|c} { 5 601 338 650 $\pm$ 50} \\
\hline\textbf{$5s^2\,^{1}S_0-5s5p\,^3P_0$} &\\
 \hline indirect measurement& \multicolumn{1}{|c} {429 228 004 300$\pm$110} \\
\hline direct measurement& \multicolumn{1}{|c} {429 228 004 235 $\pm$20} \\
\hline\hline
\end{tabular}
\end{center}
\label{tab:Freq}
\end{table}

The technique used for the direct detection of this line can be
extended to other atomic species. As an example, Yb has the same
level scheme as Sr with several isotopes having non-zero nuclear
spin. With Yb atoms trapped on the $^1S_0-^3P_1$ transition of
natural width 200 kHz, the lifetime of the MOT can be much longer
than the blue Sr MOT, up to several seconds in Ref.\
\cite{Kuwamoto99}. The contrast of the forbidden line would then
be close to 100\%.

The observation of the $^1S_0-\,^3P_0$ transition is a first step
towards the realization of an optical frequency standard using
trapped neutral atoms. Clearly the next step is the measurement of
the wavelength of the dipole trap beam where light-shift
cancellation occurs. This measurement requires better frequency
resolution than achieved here, i.e. higher laser intensity at 698
nm and/or colder atoms.

In this new type of optical frequency standards a line-$Q$ as high
as $10^{15}$ is achievable if the spontaneous emission rate in the
optical trap is less than one per second. With a laser intensity
of $2\times 10^7$ W/m$^2$, the trap oscillation frequency can be
50 kHz and the spontaneous emission rate 0.6 s$^{-1}$. With a
line-$Q$ of $10^{15}$ and a reasonable trapped atom number of the
order of $10^6$, ultimate performances are orders of magnitude
better than existing devices. The frequency noise of the laser
used to probe the atoms will then be of decisive importance
\cite{Quessada03}. State of the art ultra stable lasers use
macroscopic resonators as a reference and exhibit $1/f$ frequency
noise at Fourier frequencies below a few Hz \cite{Young99}. We
propose to circumvent this problem with a first stage servo
control to the atomic transition with a separate setup fully
optimized for frequency stability: large number of atoms,
favorable duty cycle\ldots At the price of an increase of the
experimental width of the atomic resonance the response time of
the servo-control can be as fast as desired. With this additional
degree of freedom, the optimization of ultra-stable lasers should
lead to a large improvement of their performances. One could then
approach the demanding requirements of the optical standard using
trapped neutral atoms.

We thank Ouali Acef, Andr\'{e} Clairon, Michel Lours and Giorgio
Santarelli for helpful discussions, and the optoelectronic group
of the university of Bath (UK) for providing the photonic cristal
fiber of the frequency chain. A. B. acknowledges his grant from
the european Research Training Network CAUAC. BNM-SYRTE is Unit\'e
Associ\'ee au CNRS (UMR 8630).


\begin{thebibliography}{10}
\bibitem{Note} We also performed absolute frequency measurements of $^1S_0-\,^3P_1$,
$^3P_0-\,^3S_1$ and $^3P_1-\,^3S_1$ transitions for all hyperfine
sublevels of $^{87}$Sr and $^{88}$Sr. These measurements will be
reported subsequently.
\end{thebibliography}

\end{document}